\documentclass{emulateapj}

\usepackage{graphics,graphicx,natbib}

\newcommand{\msun}      {\mbox{M$_{\odot}\,$}}
\newcommand{\lsun}      {\mbox{L$_{\odot}\,$}}
\newcommand{\kms}       {km~s$^{-1}\,$}

\slugcomment{Accepted by the Astrophysical Journal, March 7, 2007}

\shortauthors{McCRADY \& GRAHAM}
\shorttitle{SUPER--STAR CLUSTERS IN M82}

\begin{document}

\title{ SUPER--STAR CLUSTER VELOCITY DISPERSIONS AND VIRIAL MASSES IN
THE M82 NUCLEAR STARBURST\altaffilmark{1} }

\author{ Nate McCrady\altaffilmark{2,3} and James R. Graham} 

\affil{Department of Astronomy, University of California, Berkeley, CA
94720-3411}

\altaffiltext{1}{ Based on observations made at the W.M. Keck
Observatory, which is operated as a scientific partnership among the
California Institute of Technology, the University of California and
the National Aeronautics and Space Administration.  The Observatory
was made possible by the generous financial support of the W.M. Keck
Foundation.}

\altaffiltext{2}{
Now at Department of Physics and Astronomy, UCLA, Los Angeles, CA 90095-1547}

\altaffiltext{3}{
nate@astro.ucla.edu}


\begin{abstract} 

We use high-resolution near-infrared spectroscopy from Keck
Observatory to measure the stellar velocity dispersions of 19 super
star clusters (SSCs) in the nuclear starburst of M82.  The clusters
have ages on the order of 10 Myr, which is many times longer than the
crossing times implied by their velocity dispersions and radii.  We
therefore apply the Virial Theorem to derive the kinematic mass for 15
of the SSCs.  The SSCs have masses of $2 \times 10^5$ to $4 \times
10^6$ \msun, with a total population mass of $1.4 \times 10^7$ \msun.
Comparison of the loci of the young M82 SSCs and old Milky Way
globular clusters in a plot of radius versus velocity dispersion
suggests that the SSCs are a population of potential globular clusters.
We present the mass function for the SSCs, and find a power law fit
with an index of $\gamma = -1.91 \pm 0.06$.  This result is nearly
identical to the mass function of young SSCs in the Antennae galaxies.

\end{abstract}



\section{Introduction}

\subsection{Starburst Galaxies and Super Star Clusters}
\label{intro_sscs}

Short-duration episodes of intense star formation known as
``starbursts'' are responsible for a significant portion of star
formation activity in the present-day Universe.  \citet{heckman98}
estimates that the four most luminous circumnuclear starbursts (M82,
NGC 253, M83 and NGC 4945) account for 25 percent of the high-mass
($>$ 8 \msun) star formation within 10 Mpc.  The starburst phenomenon
is the present-day manifestation of the dominant mode of star
formation in the early Universe \citep{leitherer01}.  At $z=0$,
high-mass stars form predominantly in dense clusters and OB
associations \citep{miller78}.  Massive stellar clusters in nearby
starburst galaxies thus provide a laboratory for studying intense star
formation and related feedback processes, as well as physical
conditions analogous to high-redshift star formation.

Star formation in starbursts is resolved into young, dense, massive
``super--star clusters'' (SSCs) that represent a substantial fraction
of new stars formed in a burst event \citep{meurer95,zepf99}.  {\it
Hubble Space Telescope} (HST) observations with WFPC/WFPC2 in visible
light \citep[e.g.,][]{o'connell94,whitmore95} have resolved SSCs in
the nearest starburst galaxies, and SSCs appear ubiquitous in mergers
and interacting galaxies.  Of the roughly 30 gas-rich mergers observed
by HST, all have young, massive, compact clusters \citep[][and
refs.]{whitmore01}.  A spectacular example is the ``Antennae''
galaxies, NGC 4038/4039, with more than $10^3$ optically-visible SSCs
\citep{whitmore99} and many other clusters deeply embedded in dust
\citep{gilbert00}.

The derived masses, radii and ages of SSCs suggest that they are young
globular clusters, and the brightest and most massive may evolve into
a population of globular clusters similar to that of the Milky Way.
The question of whether SSCs are in fact the progenitors of globular
clusters depends critically on their masses and their content of
low-mass stars in particular \citep{meurer95}. If the stellar initial
mass function (IMF) within the clusters is biased toward high-mass
stars (or ``top heavy''), for example by suppression of low mass star
formation, the clusters may not survive the disruptive nature of mass
loss resulting from both stellar evolution and dynamical processes
\citep[e.g.][]{chernoff90,takahashi00}.

Local analogues of SSCs --- massive, dense young clusters in the
Galaxy and Large Magellanic Cloud --- contain substantial populations
of low-mass stars.  The central ionizing star cluster in NGC 3603, the
most massive H~II region in the Galaxy, has 2000 \msun in OB stars
alone. \citet{brandl99b} find that the mass spectrum of the cluster is
``well populated'' down to 0.1 \msun.  The cluster R136 at the center
of the 30 Doradus nebula in the Large Magellanic Cloud is considered
the closest example of a starburst region \citep{brandl96}.  In
addition to $\sim 10^3$ OB stars, some with masses in excess of 100
\msun, \citet{sirianni00} find a ``substantial population'' of
low-mass stars down to masses of 0.6 \msun in R136.  With a cluster
mass of $\sim 2 \times 10^4$ \msun \citep{walborn02}, R136 is at the
lower end of the SSC mass range.

\subsection{Mass Estimates}

Most mass estimates to date for SSCs beyond the Local Group are
``photometric masses.''  This technique involves measuring the
luminosity and broadband color of a cluster, and comparing the results
to the predictions of spectral synthesis models.  Examples include
optical studies of clusters in the Antennae galaxies \citep[][and
refs.]{whitmore00} and the nuclear cluster in IC 4449 \citep{boker01}.
The resulting values are highly dependent on the assumed IMF, age
estimates and theoretical stellar evolution models.  Consequently, the
photometric method provides only very weak constraints on the IMF in a
cluster.  The question of how SSCs evolve or whether they can survive
to become globular clusters cannot be directly addressed with
confidence by photometry alone.

Recently, however, several studies have obtained kinematic masses for
SSCs in starbursts directly from observations of stellar velocity
dispersions.  \citet{hoflipper96b} use high-resolution optical spectra
to measure the velocity dispersion of an SSC in the nearby amorphous
galaxy NGC 1705.  They derive a cluster mass of $(8.2 \pm 2.1) \times
10^4$ \msun.  The dwarf starburst galaxy NGC 1569 contains two
prominent SSCs.  \citet{hoflipper96a} used optical spectroscopy to
determine the velocity dispersion and derive a mass of $(2-6) \times
10^5$ \msun for SSC-A.  \citet{gilbert02th} uses near-IR spectroscopy
to identify two separate velocity components along the line of sight
at the position of SSC-A, and finds masses of $3.0 \times 10^5$ \msun
for cluster A1 and $3.4 \times 10^5$ \msun for cluster A2.  Gilbert
also finds a velocity dispersion for SSC-B and derives a mass of $1.8
\times 10^5$ \msun.  \citet{mengel02} use high-resolution optical and
near-IR spectroscopy to measure the velocity dispersions of six of the
brightest clusters in the merging Antennae galaxies.  They derive mass
estimates ranging from $6.5 \times 10^5$ to $4.7 \times 10^6$ \msun.

In addition to these clusters in starburst galaxies, the kinematic
approach has been applied to several young, massive clusters in more
quiescent galaxies.  \citet{boker99} uses near-IR spectra to derive a
mass of $6.6 \times 10^6$ \msun for the nuclear star cluster in the
giant spiral IC 342.  \citet{larsen04} derives virial masses for two
SSCs in the dwarf irregular NGC 4214 and two SSCs in NGC 4449 based on
optical spectra, and one young SSC in the nearby spiral NGC 6946 based
on near-IR spectra.  Clusters in the Larsen study have masses between
$2 \times 10^5$ and $1.8 \times 10^6$ \msun, typical of SSCs.


Mass estimates based on measured velocity dispersions are nearly
independent of theoretical models, relying on simple application of
the virial theorem.  Armed with the virial mass, one can derive the
light-to-mass ratio of an SSC, which may be compared to predictions of
population synthesis models to constrain the cluster's IMF.  Critical
to understanding the IMF is detection of low-mass stars, the light of
which is swamped by high luminosity supergiants.  Measurement of the
kinematic mass provides the only means of detecting and quantifying
the low-mass stellar population of a cluster based on its integrated
light.

\subsection{Messier 82}
\label{intro_m82}

M82 (NGC 3034) provides a useful laboratory.  As one of the nearest
starburst galaxies at 3.6 Mpc \citep{freedman94}, M82 presents an
obvious resolution advantage: star-forming regions can be studied on
small spatial scales ($1'' = 17.5$ pc) and individual SSCs are
resolved by HST.  The galaxy's high inclination of 81$^\circ$
\citep{achtermann95} and prevalent dust lead to large, patchy
extinction; infrared observations are required to overcome this
obstacle in characterizing the SSC population.

\begin{figure}
\centering
\epsscale{1.0}
\plotone{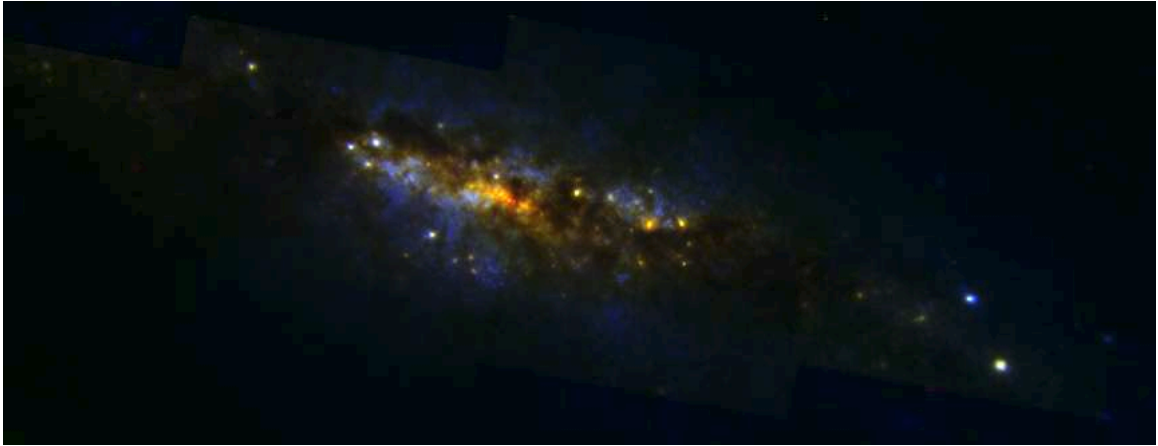}
\caption
{Color mosaic of HST ACS/WFC and NICMOS images of the nuclear region
in M82.  ACS F814W, NICMOS F160W and NICMOS F222M images are mapped to
blue, green and red, respectively.  The image is $\sim$ $25'' \times
65''$ ($0.4 \times 1.1$ kpc) with north up and east to the left.
About two dozen super star clusters are evident, many of which are
spatially coincident with and reddened by the band of variable
extinction running from upper left to lower right in the image.}
\label{m82mosaic}
\end{figure}

The light of red supergiant stars (RSGs) dominates the near-IR
continuum throughout the galaxy's starburst core.  \citet{satyapal97}
interpret the compact sources along the plane of M82 as young star
clusters, the space density of which increase towards the nucleus.
The smooth component of the near-IR emission is itself likely the
integrated contribution from unresolved clusters of RSGs
\citep{natascha98}.  Based on mid-IR observations, \citet{lipscy04}
find that at least 20 percent of the star formation in M82 is
occurring in SSCs.  HST/NICMOS images of the region (Figure
\ref{m82mosaic}) show many luminous SSCs within $\sim 300$ pc of the
nucleus.

The nuclear starburst is ``active'' in the sense that the typical age
for the starburst clusters is $\sim 10^7$ years \citep{satyapal97}.
Evolutionary synthesis models by \citet{natascha98} suggest the
nuclear starburst (i.e., the central 450 pc) consists of two distinct,
short duration events with ages of about 5 Myr and 10 Myr.  The most
intense star formation took place parallel to the plane of the galaxy
with a peak near the nucleus.  \citet{o'connell95} image M82 in the
$V$ and $I$ bands with the high-resolution Planetary Camera aboard
HST, identifying over 100 SSCs within a few hundred parsecs of the
nucleus.  \citet{degrijs01} images a region in the disk of M82, 1 kpc
from the nuclear starburst, with WFPC2 and NICMOS.  They identify 113
SSC candidates which were part of a starburst episode $\sim$ 600 Myr
ago (a ``fossil starburst''), with little star formation in the past
300 Myr.  The clusters in the fossil starburst have masses of
$10^{4-6}$~\msun.  \citet[][SG01]{smith01} estimate an age of $60 \pm
20$ Myr for the SSC `M82-F', intermediate between the ongoing nuclear
burst and the fossil burst farther out in the disk.  M82-F lies $\sim$
500 pc west of the nucleus of M82.  \citet{mccrady05} measures a mass
of $5.6 \times 10^5$ \msun for the cluster based on near-IR
observations, and found evidence for mass segregation.

Early ground-based studies of M82 found evidence of an abnormal IMF.
\citet{rieke93} uses population synthesis models to constrain the IMF
based on the near-IR observations of \citet{mcleod93}.  They conclude
that the large $K$-band luminosity of the M82 starburst relative to
its dynamical mass requires an IMF that is significantly deficient in
low-mass stars ($M < 3$ \msun).  \citet{doane93} examine the
supernova rate, molecular gas mass and total dynamic mass and conclude
that an IMF producing stars of mass $> 3$ \msun easily matches
observations, whereas a power-law IMF \citep[e.g.,][]{salpeter55}
would require an unreasonably small mass of stars in the region prior
to the onset of the burst.  In contrast to these global studies,
\citet{satyapal97} use $1''$-resolution near-IR images to identify
pointlike sources and find that at this scale starburst models can
match observations without invoking a high-mass-biased IMF.  High
spatial-resolution studies are necessary to investigate star formation
in the cluster-rich M82 starburst.  In a pilot study for this article,
\citet{mccrady03} measure the kinematic mass of two clusters using
near-IR spectra and imaging.  Based on the light-to-mass ratios of the
clusters, they find that one (MGG-11) appears to have a top-heavy IMF,
whereas the other (MGG-9) appears consistent with a normal IMF.
Measurement of the SSC mass independent of assumptions regarding the
$L/M$ ratio are required to further constrain the stellar IMF in the
clusters.

\subsection{Overview}

In this article, we measure the virial masses of the super star
cluster population of the inner $\sim 500$ pc of the M82 starburst,
extending the work of \citet{mccrady03}.  Our aim is to examine star
formation in the starburst on the scale of individual super star
clusters, regions only a few parsecs in extent, with an eye towards
placing constraints on the IMF of individual SSCs.

We use high-spectral-resolution near-IR spectroscopy from the
W.M. Keck Observatory to measure the stellar velocity dispersions and
dominant stellar spectral type of the SSCs.  We then apply the virial
theorem to derive their masses.  Clusters for which the age may be
determined facilitate investigation of the IMF.  In \S~\ref{method} we
describe the kinematic mass, the method we use to measure the velocity
dispersion of stars in a cluster, and discuss related systematic
effects.  In \S \ref{nirspec} we discuss the NIRSPEC observations,
data reduction and spectral extraction.  In \S~\ref{dispersions} we
measure cluster velocity dispersions and derive the kinematic masses
of the clusters, and present the cluster mass function for the nuclear
starburst region.

\section{Approach} 
\label{method}

\begin{figure}
\centering
\epsscale{1.0}
\plotone{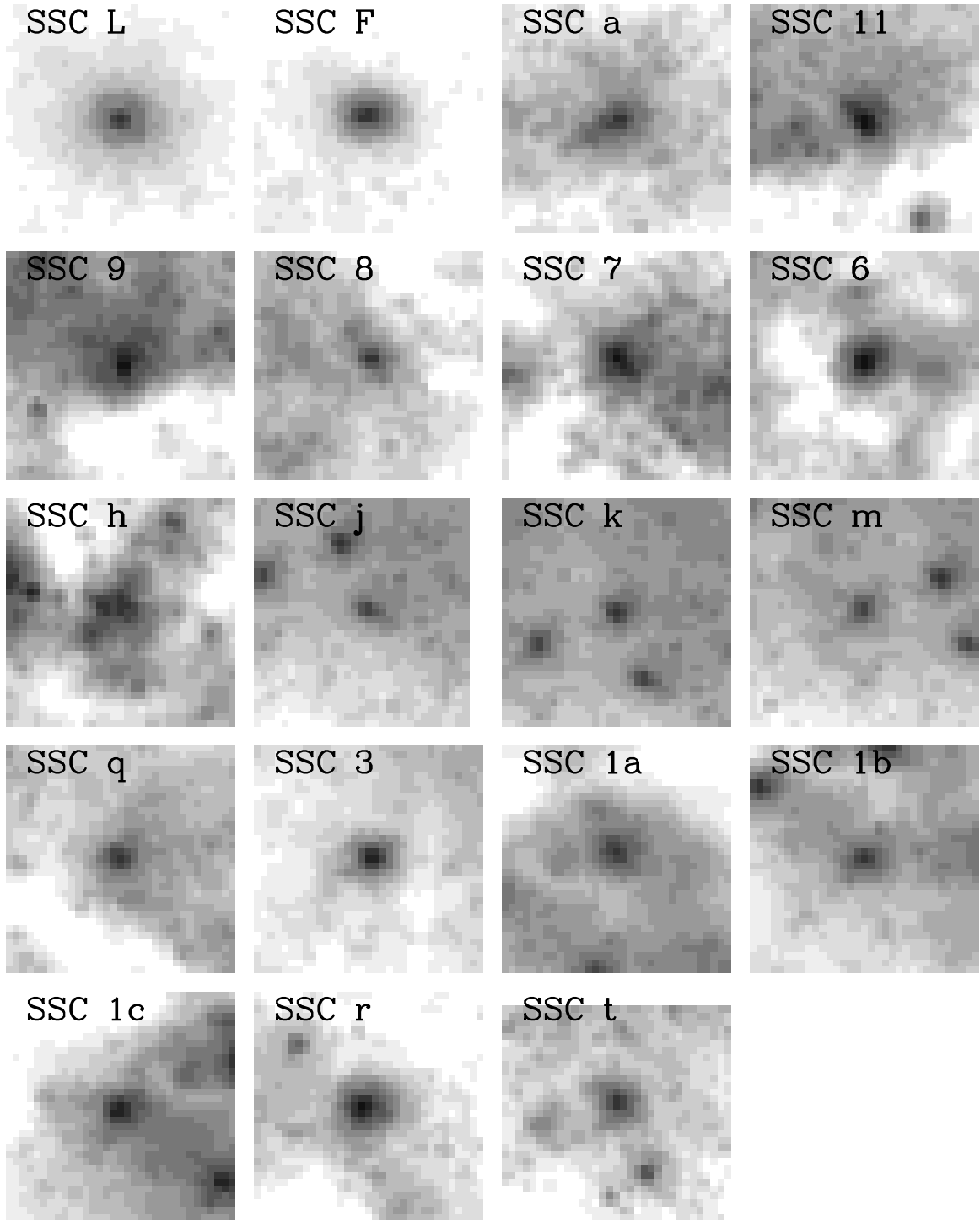}
\caption
{HST/NICMOS F160W images of each cluster.  Each image is $2.5'' \times
2.5''$, and the position angle of the y-axes is $349.4^{\circ}$
(i.e., North is $10.6^{\circ}$ left of straight up).  The images are
log-scaled, as the cores are substantially brighter than the halos.}
\label{yearbook}
\end{figure}

The mass of a gravitationally-bound star cluster may be determined by
application of the virial theorem \citep{spitzer87}.  Specifically,
the virial mass is a function of two observable quantities:

\begin{equation}
M = 10\,\frac{r_{hp} \sigma_r^2}{G}
\label{virial}
\end{equation}

\noindent where $r_{hp}$ is the half-light radius in projection,
$\sigma_r$ is the one-dimensional line-of-sight velocity dispersion
and $G$ is Newton's gravitational constant.  Half-light radii for the
M82 clusters were measured in \citet{mccrady03} based on HST/NICMOS
images.  We assume the light profile of the cluster traces the mass
distribution, and thus use the measured half-light radius as a proxy
for the half-mass radius.  In the case of mass segregation, however,
this assumption breaks down for near-IR light, and the resulting mass
represents a lower-limit \citep{mccrady05}.  The M82 nuclear SSCs are
labelled in a NIRSPEC slit-viewing camera (SCAM) mosaic shown in
Figure \ref{sscmap}.  HST/NICMOS $H$-band (F160W) images of the
clusters are shown in Figure \ref{yearbook}.

\begin{figure}[!th]
\centering \epsscale{1.0}
\plotone{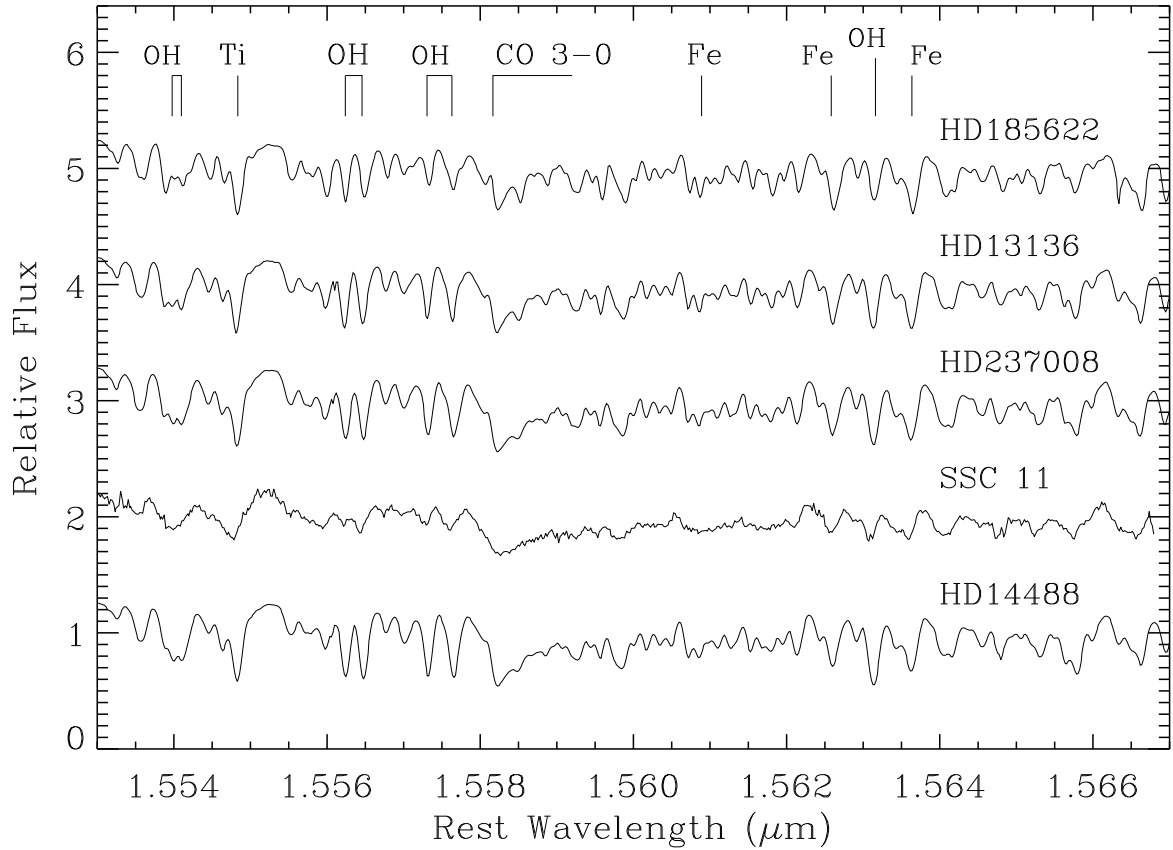}
\caption[Comparison of SSC-11 and supergiant spectra]
{Comparison of the spectra of SSC-11 and several cool supergiants in
echelle order 49.  The cluster spectrum displays the same features as
the supergiants, but appears washed out due to the velocity dispersion
of its constituent stars.  The supergiant stars are plotted in a
temperature sequence, with the hottest star at the top.}
\label{ssc11_comp}
\end{figure}

To measure $\sigma_r$ for the clusters, we obtain
high-spectral-resolution near-IR integrated light spectra and perform
a cross-correlation analysis with template supergiant stars.  The
near-IR spectrum of a young SSC (i.e., ages $< 100$ Myr) is dominated
by the light of cool, evolved supergiant stars.  These highly luminous
stars have a large number of molecular and atomic features in the $H$
band.  The integrated-light spectrum of an SSC resembles the spectrum
of a red supergiant star, the features of which have been ``washed
out'' by the velocity dispersion of stars in the cluster (Figure
\ref{ssc11_comp}).  Our cross correlation method, described in detail
in \citet{mccrady03}, returns both the velocity dispersion of the
cluster relative to a particular template supergiant and a measure of
the similarity of the cluster and supergiant as quantified by the peak
value of the cross correlation function (CCF).  We have prepared an
atlas of 19 high-resolution ($R \sim 22,000$) template star spectra in
the $H$ band, ranging from spectral types G2 through M5 in luminosity
class I \citep{kirian06}.  Results of the cross correlation analysis
are presented in \S \ref{dispersions}.

Determination of the velocity dispersion by cross correlation analysis
is subject to several potential sources of systematic error.  A
detailed analysis is presented in \citet{mccrady05th}.  We present an
overview in the following paragraphs.

One potential difficulty is metallicity differences between the
Galactic supergiants used as templates and the supergiants producing
the cluster light.  \citet[][and refs.]{mcleod93} cited evidence from
emission-line studies of various elements and concluded that the
present-day ISM in M82 has solar or slightly higher metallicity.
\citet{natascha01} determined that near-IR stellar absorption features
observed in the starburst core are consistent with the light from
solar-metallicity red supergiants (RSG).  \citet{origlia04} performed
abundance analysis on the nuclear starburst region using spectral
synthesis models for near-IR absorption and X-ray emission.  They
found an iron abundance roughly half of the solar value, but
enhancement of $\alpha$-elements to solar or slightly higher levels.
This pattern is consistent with enrichment by recursive bursts of Type
II supernovae.  The template supergiants used in our analysis also
have roughly solar metallicity, and thus we expect that metallicity
effects are unlikely to significantly bias the measured cluster
velocity dispersions.

Cross-correlation with a mismatched template spectrum can introduce
systematic bias to the velocity dispersion determination.  Tests with
supergiant spectra broadened with a Gaussian to simulate the effect of
a cluster velocity distribution indicate that the cross correlation
analysis correctly identifies the ``best fit'' based upon the peak
value of the CCF.  Increasing the level of added noise decreases the
CCF peak, but does not lead to misidentification of the best template.
We have elected to cross-correlate the cluster spectra with spectra of
single RSG stars because the light of a young coeval cluster should be
dominated by the light of the most massive stars.  At an age of $\sim
10^7$ years, this would be the light of evolved massive stars, i.e.,
the RSG stars.  Mixing a composite spectrum (with the inclusion of
intermediate mass stars) would generate a better match to the overall
spectrum (particularly the depth of absorption features --- see
below).  But for our purposes, it is more important to be able to
isolate the width of the lines resulting from the velocity dispersion
of cluster stars.

In addition to the dominant light of the RSG stars, we expect the
cluster spectra to contain a substantial contribution from
intermediate mass stars still on the main sequence.  In the $H$ band,
the spectra of A and late-B stars (with masses $\sim$ 2--6 \msun) are
essentially featureless, with the exception of the prominent (and
wide) hydrogen Brackett series absorption lines.  The prevalent
nebular emission in the disk of M82 requires us to avoid the
wavelength ranges of these hydrogen lines in our analysis.  Over the
rest of the spectral range in our analysis, an admixture of
intermediate mass star spectra would only change the slope of the
spectrum (as the near-IR spectra of these stars are essentially
thermal).  One step in our analysis is the removal of any continuum
slope, as the cross-correlation technique is used to measure the
velocity dispersion, information which is contained in the width of
the absorption lines, not in the depth of the lines or in the
continuum.  As such, addition of intermediate mass star spectra would
not affect the measured velocity dispersions or virial masses.

Filtering of the spectra in Fourier space is a source of systematic
error.  Extracted spectra (\S \ref{nirspec}) are cross-correlated with
the spectra of template supergiant stars.  The spectra are
baseline-subtracted, apodized and both high- and low-pass filtered in
Fourier space.  At the high-frequency end, the cross correlation
result is affected by random noise; high amplitude noise residuals
from sky emission line subtraction are particularly noxious, as the
effects are unpredictable and merely serve to increase uncertainty.
Low frequencies contain information about spectrum-wide residual
variations after baseline subtraction.  One likely source of such a
variation is the presence of light from intermediate-mass main
sequence stars in the integrated cluster light.  Very broad hydrogen
absorption features typical of the otherwise largely featureless
$H$-band spectra of A0V stars \citep{meyer98}, for example, would not
be removed by the low-order baseline subtraction.  Information
pertaining to the velocity dispersion of the cluster resides in the
frequencies between the extremes.

The frequency filtering applied to the NIRSPEC data leads to a
systematic error of $0-3$ \kms, which varies between echelle orders.
The offsetting correction applied to the results respresents a
correction of no more than 20 percent, generally less.  Noise in the
input spectrum leads to uncertainty in the correction factor in the
range of $0.1-0.5$ \kms, setting the lower bound on the precision of
the velocity dispersion measurements.

\section{Observations}
\label{nirspec}

\subsection{NIRSPEC Spectra}

The spectra used to determine the internal velocity dispersions of the
SSCs and template stars were taken with the facility near-infrared
echelle spectrometer NIRSPEC \citep{mclean98} on the 10-m Keck II
telescope on Mauna Kea, Hawaii.  We used NIRSPEC in the echelle mode,
which yields spectral resolution of $R \sim 22,000$.

The integrated light of super star clusters aged 5 to 80 Myr is
dominated by evolved supergiant and bright giant stars
\citep{gilbert02th}.  The near-IR spectrum of cool evolved stars is
replete with atomic and molecular absorption features --- no
``continuum'' in the traditional sense (i.e., a Planck thermal
spectrum) is evident.  Both the $H$ and $K$ bands offer a large number
of features that the cross correlation analysis effectively averages
over in determination of the mean feature width.  There is perhaps an
advantage to the $H$ band, in that warm circumstellar dust may veil
features in the $K$ band.  The NIRSPEC detector experiences
significant ``persistence'' from exposure to large flux, for example
bright sky OH emission lines or arc lamp lines.  Operationally, this
discourages changing of filters during an observing night as the
persistent after-images of sky emission lines from a different filter
add significantly to the noise in an echellegram.  In this analysis,
we have opted to observe the clusters at the shorter wavelength only.

The NIRSPEC-5 (N5) order-sorting filter covers the wavelength range
1.51--1.75 $\mu$m, corresponding approximately to Johnson $H$.  The N5
echelle data fall in seven echelle orders, ranging from 44 through 50.
All observations were taken with the echelle and cross-dispersion
gratings set at their blaze angles.  This position maximizes
signal-to-noise for a given exposure time.  This advantage is
mitigated by the fact that more than a single position is required to
cover the free spectral range at 1.6 $\mu$m, and portions of the $H$
band are not observed.

Spectra used in this work were obtained over four observing seasons,
from February 2002 through January 2005.  Table \ref{obs1} presents a
summary list of spectroscopic observations.  NIRSPEC observations of
evolved stars used as template spectra are discussed in
\citet{kirian06} and \citet{mccrady03}.  The minimum airmass of M82
(declination $+69^{\circ}40'$) from Mauna Kea is 1.56, and efforts
were made to observe at an airmass of no larger than 2.0 when
possible.  The slit used has a width of $0.432''$ (3 pixels) and
length of $24''$.  Use of the long slit improves background
subtraction, and often allows multiple clusters to be observed
simultaneously.  Slit positions were chosen to include multiple
objects where possible to increase observing efficiency.  Certain
pairs of targets are closely-separated and only resolvable in good
seeing.

Each individual cluster spectrum has an integration time of 600
seconds.  Bright OH sky emission lines begin to saturate in longer
exposures, increasing the difficulty of sky subtraction.  Total
integration time on a cluster is increased by repeating the
observations.

\subsection{Reduction and Extraction}
\label{extract}

The spectra were dark-subtracted, flat-fielded and corrected for
cosmic rays and bad pixels.  The curved echelle orders were then
rectified onto an orthogonal slit-position versus wavelength grid
based on a wavelength solution from sky (OH) emission lines.  Each
pixel in the grid has a width of $\delta\lambda = 0.019$ nm.  We 
sky-subtracted by fitting third-order polynomials to the 2D spectra
column-by-column.

The NIRSPEC echelle turret is jostled when the cryogenic image rotator
undergoes large slews, leading to stochastic shifts of the wavelength
scale of up to several pixels.  Doppler shift information is lost due
to such shifts.  Typically, the absolute wavelength scale is
established using telluric OH emission.  Throughout each data
acquisition cycle the image rotator was either turned off, or only
slow, small amplitude tracking motions were executed. In either case
the wavelength solution is stable to better than a few hundredths of a
pixel.  Thus the velocity broadening reported here is intrinsic to the
source and is not an instrumental effect.

The cluster spectra were extracted using Gaussian weighting functions
matched to the wavelength-integrated profile of each cluster.  To
correct for atmospheric absorption in the cluster spectra, we observed
a hot main sequence star at a similar airmass.  This calibration star
spectrum is divided by a spline function fit to remove photospheric
absorption features (particularly Brackett series and helium lines)
and continuum slope.  The resulting atmospheric absorption spectrum is
then divided into the cluster spectra.

The adopted sky-subtraction method generally completely removes sky
emission lines.  However, a high level of noise is left behind at the
position of bright OH lines, particularly un- or barely-resolved
doublets.  These ``noise spikes'' must be removed to avoid
introduction of systematic bias in the cross-correlation results.  We 
smooth the spectrum with a broad ($\sim 40$ \kms) step function, and
subtract the original spectrum to obtain a residuals array.  Data
points greater than $5\times$ the rms are replaced by the smoothed
pixel.  The step-function width and clipping level were chosen to
limit replaced pixels to only those affected by strong sky emission
lines.  Certain atmospheric OH emission lines were incompletely
removed in the sky subtraction process.  In these cases, we replaced
the pixels affected by residual sky emission with the median value of
the $\sim 5$ pixels on either side of the contaminated range.  The
fraction of pixels replaced in a given echelle order typically amounts
to a few percent.  Tests on an OH emission-free echelle order indicate
that the cross-correlation result is unaffected within the stated
uncertainties.

\begin{figure}
\begin{center}
\epsscale{.95}
\plotone{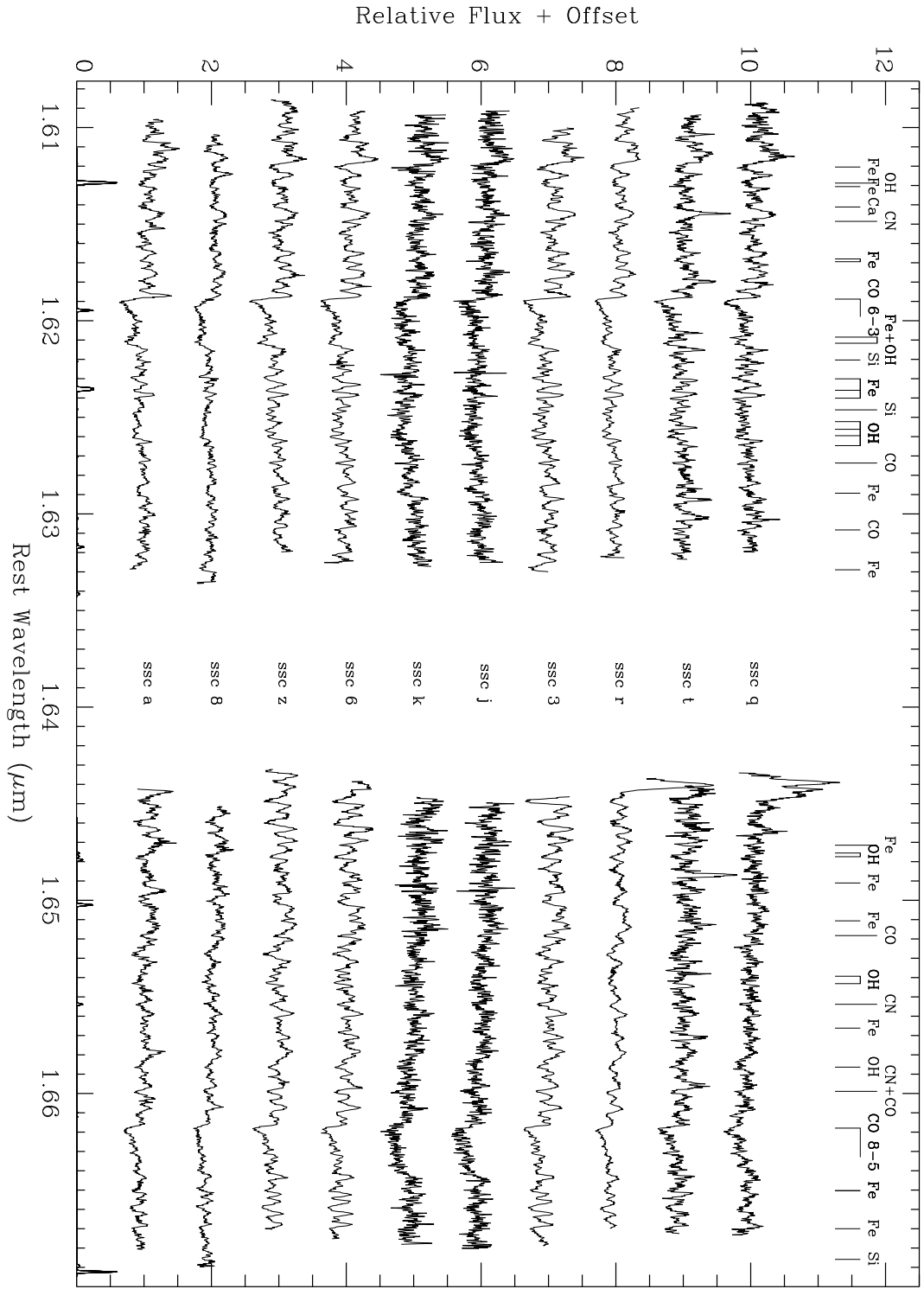}
\caption{Atlas of SSC spectra for echelle orders 47 \& 46.  }
\label{atlas1}
\end{center}
\end{figure}

\begin{figure}
\begin{center}
\epsscale{.95}
\plotone{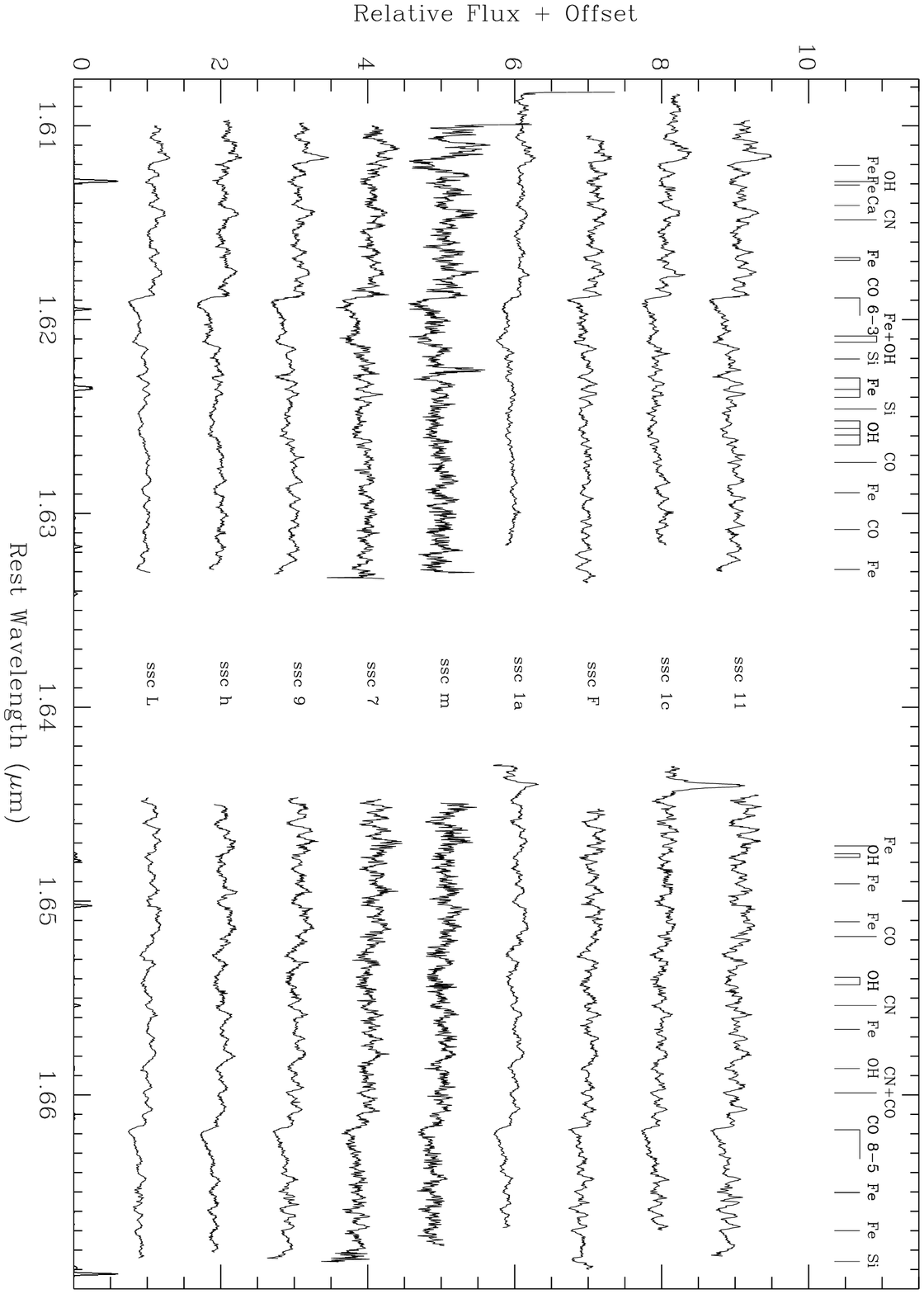}
\caption{Atlas of SSC spectra for echelle orders 47 \& 46, continued.  }
\label{atlas2}
\end{center}
\end{figure}

An atlas of the spectra for 19 SSCs is presented in Figures
\ref{atlas1} and \ref{atlas2}.  We have included here only the spectra
for echelle orders 46 and 47; plots of all echelle orders for each
cluster are available in \citet{mccrady05th}.  Each cluster spectrum
represents the summation of multiple observations (Table \ref{obs1}).
The total is normalized by dividing by the median value for that
echelle order, such that the spectrum is centered about unity.  The
resulting scale is relative flux, which allows direct comparison of
spectra of different clusters.  The spectra are offset by an arbitrary
integer amount for presentation of multiple clusters on a single set
of axes, and labeled towards the right (long-wavelength) side.  The
clusters within a given echelle order are arranged in the atlas in
order of increasing velocity dispersion.

The signal-to-noise ratio ($S/N$) of the extracted spectra varies
between clusters.  The clusters vary in luminosity over a range of
four apparent magnitudes in $H$ band \citep{mccrady03}, which is a
factor of $\sim 40$.  The luminosity differences carry through to the
single-integration $S/N$ as all clusters were observed for 600 seconds
per integration.  Light losses due to variable seeing and
inefficiencies in fine guiding add additional variation to the $S/N$
for each cluster.  Although faint clusters were observed more often,
not all clusters were observed to the same $S/N$ as a result of
observing constraints.  Differences in the total $S/N$ are evident in
the atlas of cluster spectra.  For example, faint SSC-k was observed
only twice and has total $S/N \sim 9$ per pixel based on the CCD
equation \citep[][ p. 54]{howell00}, which accounts for Poisson
statistics, background, dark current and read noise.  Bright SSC-1c
was also observed just twice, but has total $S/N \sim 28$ per pixel.
Repeated observations (13 times) of SSC-r brought the total $S/N$ up
to $\sim 37$ per pixel; a single observation of faint SSC-r results in
$S/N$ comparable to a single observation of SSC~k.  These examples
provide a sense of the S/N range of the observations.

The spectra presented in the atlas have been shifted to rest
wavelength.  Across the top of each plot, we have identified the
positions of certain prominent spectral features.  The $H$-band
spectra of the clusters resemble the spectra of supergiant stars, and
have a large number of iron and OH absorption lines.  Rovibrational
$\Delta v = 3$ bandheads of carbon monoxide are recognizable in orders
45 through 49.  As seen in the spectra of supergiant stars
\citep{kirian06,meyer98}, the strength of the CO bandheads and OH
lines increase at cooler effective temperatures.  Lines from other
miscellaneous metals (Mn, Ti, Si, Ca, C, Ni) and molecules (CN) are
also indicated.  At the bottom of each atlas plot is a representative
sky emission spectrum for that echelle order, arbitrarily scaled.  The
brightest OH emission lines leave a footprint of increased
noise in the extracted cluster spectrum. 

The disk of M82 has substantial diffuse emission.  The nuclear
starburst displays mottled near-IR continuum emission (Figure
\ref{m82mosaic}).  The cross-correlation analysis based upon the
spectra of the clusters is more sensitive to line emission.
\citet{lynds63} first noted the bipolar, filamentary network of
H$\alpha$ emission extending more than 1 kpc from the galactic center.
Paschen $\alpha$ images show patchy recombination emission throughout
the nuclear starburst \citep{alonso03}.  $K$-band spectra show
emission lines of hydrogen (Br~$\gamma$), He I and H$_2$.  Prominent
$H$-band emission lines are [Fe~II] at 1.644 $\mu$m and the Br~6 line
of hydrogen at 1.7367 $\mu$m.  In several of the slit positions used,
nebular emission varies along the slit and removal is difficult.
Remnants of the lines are apparent in the spectra of certain clusters
(e.g., SSC-1c in Figure \ref{atlas2}).

\subsection{Objects Observed}

\begin{figure}
\begin{center}
\epsscale{0.95}
\plotone{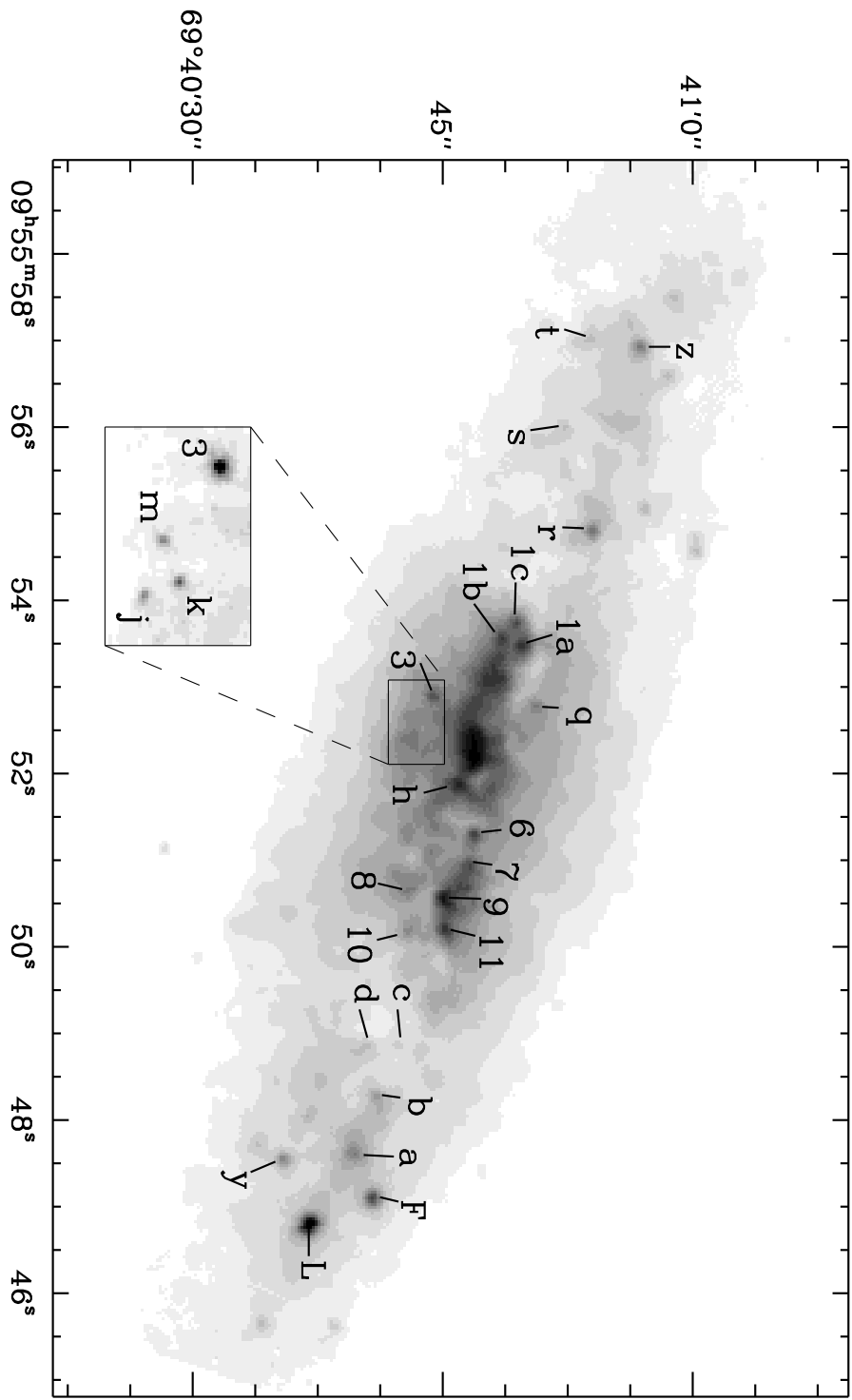}
\caption
{\label{sscmap}Mosaic of $H$-band (N5) NIRSPEC SCAM images of the
nucleus of M82.  Candidate SSCs are labeled for reference.
Coordinates are J2000.  Inset image is from HST/NICMOS.}
\end{center}
\end{figure}

The 20 objects of Table 3 of \citet{mccrady03} and the 19 objects of
Table \ref{disptable} do not constitute equal sets.  The intersection
of the two tables contains the 15 SSCs for which we have herein
derived a mass (see Table \ref{disptable}).  The union of the two
tables contains 24 objects.  Moreover, Figure \ref{sscmap} identifies
26 objects.  An accounting of the objects observed in this project is
as follows.

Five clusters for which we measured photometry and half-light radius
had significant problems with their echelle spectroscopy.  The
practical single-exposure time limit of 600 seconds is set by the
saturation point of atmospheric OH emission lines.  Clusters fainter
than [F160W] $\sim 15$ mag have $S/N < 2$ in 600 seconds of typical
seeing.  In poor seeing, these faint clusters are often too difficult
to identify for positioning the spectrograph slit.  SSC-1b is a
frustrating case, as it is a bright cluster with good $S/N$ in a 600-s
exposure.  In poor seeing, however, the light from SSC-1b is blended
with the light of the nearby clusters SSC-1a and SSC-1c.  Observations
on 2003 Feb 6 were not used for SSC-1b or SSC-1c because of inadequate
seeing.  On the night of 2005 Jan 24, the seeing was sufficient to
resolve SSC-1c, however the NIRSPEC detector was contaminated by
persistence due to sky emission from use in low-resolution mode by an
unaffiliated observing team earlier in the night.  While we were able
to extract SSC-1c, SSC-1b was significantly contaminated and had to be
rejected.

Six other objects identified in Figure \ref{sscmap} have no derived
virial mass.  SSC-z is clearly a super star cluster, but lies just
north of the edge of the HST/NICMOS field \citep{mccrady03}.  In the
absence of a resolved image, we have no measurement of the half-light
radius.  Object ``y'' also lies off the edge of the NICMOS field, just
south of SSC-L.  Spectra of object ``y'' are inconclusive due to low
$S/N$ in any case.  Object ``10'' is unresolved by the NICMOS image,
and we therefore have no measured half-light radius.  Interestingly,
object ``10'' is coincident with a point source in Paschen $\alpha$
images, suggesting it may be a compact H~II emission region
surrounding one or several massive stars.  SSC-j and SSC-a are not
well fit by empirical King functions, and the half-light radii of
these clusters are undetermined.  SSC-h is likewise not well fit by
the empirical King model, as it is clearly a collection of sources in
NIC2 images (Figure \ref{yearbook}).

\section{Analysis}
\label{dispersions}

\subsection{Cross Correlation Results}

Each cluster was observed multiple times, with seven echelle orders in
the N5 filter per observation.  A single observation of the spectrum
in a particular echelle order is treated as one ``experiment'' for the
cluster.  Each of the experiments for a cluster is cross correlated
with the spectrum from the corresponding spectral order of each of the
template evolved stars.  The result of this analysis is an ensemble of
cross correlation functions (CCFs) for each cluster/template star
pair.  The peak amplitude of the CCF measures the similarity of the
cluster to the template spectral type (Table \ref{disptable}).  For
each cluster, we have identified the template supergiant spectrum
which provides the best match.  In most cases, several template stars
match the cluster approximately equally well.  The spectra of template
stars with higher surface temperatures (e.g., G-type stars) proved to
be poor matches for the cluster spectra.  The CCFs for each cluster
and its best-match template are presented in \citet{mccrady05th}.

The velocity dispersions based upon cross correlation results for the
best-match template star are listed in Table \ref{disptable}.  The
quoted uncertainties reflect the formal error based upon the standard
deviation of the mean for the ensemble of experiments.  Systematic
errors are treated in \citet[][see also \S \ref{method}]{mccrady05th};
the stated uncertainties do not include any allowance for the applied
correction of systematic offsets.  In the course of the analysis, each
cluster spectrum is cross correlated with each of the template
spectra.  We find that the velocity dispersions indicated by the best
match template are consistent with velocity dispersions indicated by
other templates of similar effective temperature to within the
uncertainties.

\subsection{Derived Virial Masses}
\label{derivedmasses}

Armed with measurements of the cluster half-light radii and velocity
dispersions, we are ready to derive the virial masses.  Table
\ref{disptable} lists the derived mass for 15 SSCs in M82.  Most of
the SSCs have masses between $2 \times 10^5$ and $10^6$ \msun;
clusters SSC-L, SSC-7 and SSC-9 have $M > 10^6$ \msun.  The median
uncertainty on the virial mass measurements is 16 percent.  A
significant portion of the error budget is the 8 percent uncertainty
in the adopted distance to M82.

\begin{figure}
\centering 
\epsscale{1.0} 
\plotone{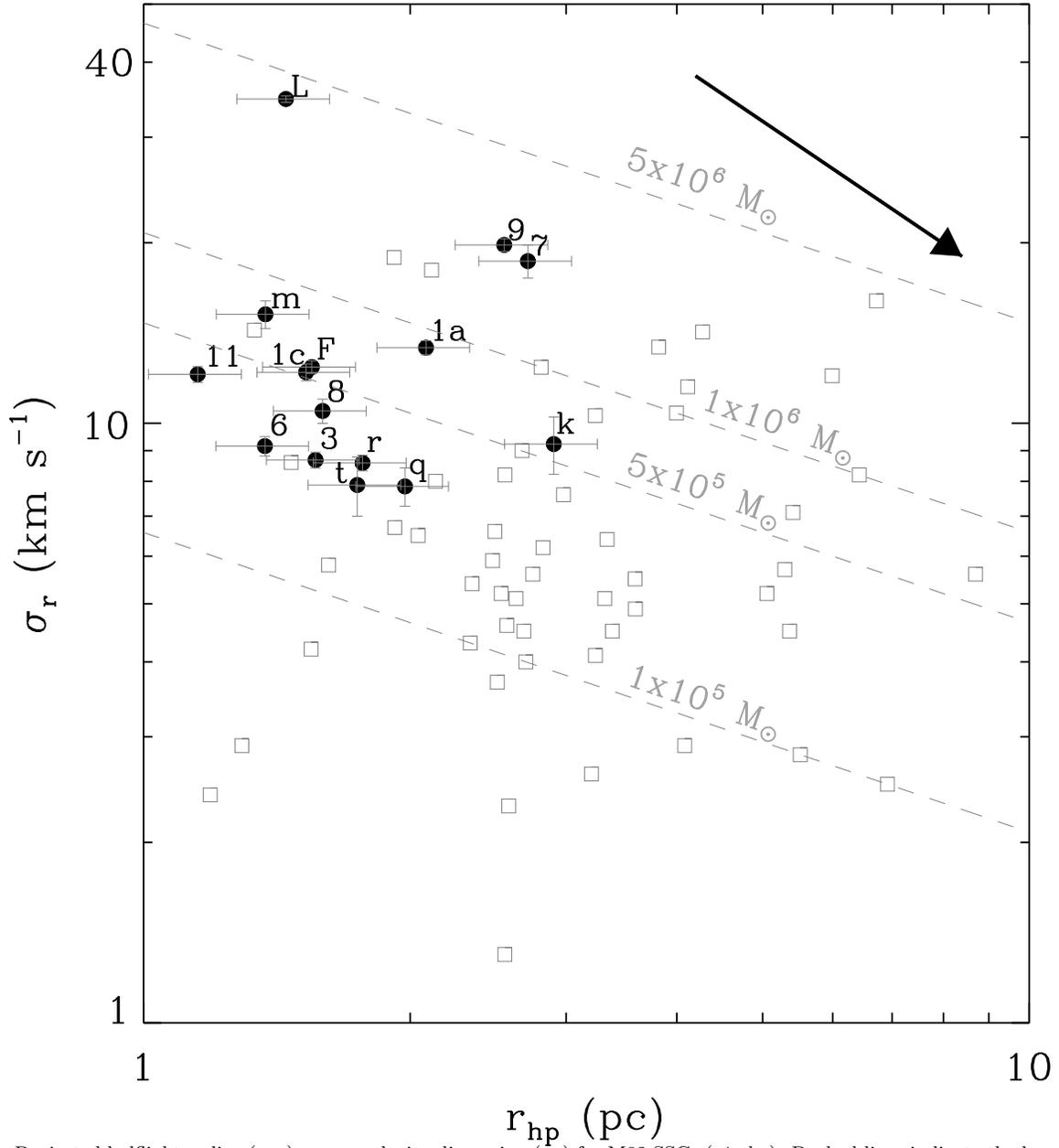}
\caption
{Projected halflight radius ($r_{hp}$) versus velocity dispersion
($\sigma_r$) for M82 SSCs ({\it circles}).  Dashed lines indicate the
locus of points for cluster mass as labeled.  Error bars on the
halflight radius do not include the uncertainty on the distance to
M82.  Galactic globular clusters ({\it squares}) from \citet{pryor93}
are plotted for comparison.  The vector indicates time evolution of a
cluster due to adiabatic loss of half its mass (see Appendix).}
\label{virialplot}
\end{figure}

Figure \ref{virialplot} plots the half-light radii versus the velocity
dispersions for the SSCs.  We have plotted the locus of points for
certain masses as dashed lines.  These lines illustrate that the
uncertainty in the velocity dispersion, $\sigma_r$, has a greater
impact on the uncertainty in mass than does the uncertainty in the
half-light radius, $r_{hp}$.  This is to be expected, as the virial
mass (Eq. \ref{virial}) is proportional to the square of $\sigma_r$.
We have mitigated this effect by measuring $\sigma_r$ to a precision
sufficient to balance the error budget roughly evenly between
uncertainties on the velocity dispersions and the halflight radii.

Application of the virial theorem to determine the mass of the
clusters is based on the assumption that the clusters are at present
bound (self-gravitating) entities.  This assumption is supported by a
comparison of the relevant timescales: the crossing time and the age
of the clusters.  The crossing time is the typical time required for a
star to cross the cluster, where $t_{cr} \approx r_{hp}/\sigma_{r}$
\citep[][p. 190]{binney87}.  The SSCs in Table \ref{disptable} have
crossing times in the range of $4 \times 10^4$ to $3 \times 10^5$
years, while their ages are on the order of $\sim 10^7$ years
\citep{satyapal97}.  Thus, member stars have made tens of crossings of
the clusters.  After just a few crossing times, the stars of a cluster
are well mixed \citep{king81} and the virial theorem is well satisfied
\citep{aarseth74}.  The assumption that the M82 clusters are currently
gravitationally bound is therefore valid.

To provide context and a sense of scale for the derived cluster
masses, we turn to virial mass measurements of SSCs and globular
clusters in our own and other galaxies from the literature.
\citet{pryor93} use velocity dispersions and King-Michie model fits to
derive virial masses of 56 Galactic globular clusters.  Their velocity
dispersions range from $1-19$ \kms ($\overline{\sigma} = 6.8$ \kms)
and masses of $10^4$ to $4 \times 10^6$ \msun ($\overline{M} = 5.6
\times 10^5$ \msun).  The Milky Way globular clusters are plotted in
Figure \ref{virialplot} for comparison with the M82 clusters.
(\citet{pryor93} provide $\sigma_r$ and $M$, from which we estimated
$r_{hp}$ using Eq.  \ref{virial}.)  In total, the Milky Way has about
180 globular clusters \citep[][p. 31]{ashman98}.  If we assume an
average cluster mass of $1.9\times 10^5$ \msun \citep{mandushev91},
the total mass of the Galactic population is $\sim 3.4 \times 10^7$
\msun.  The M82 SSCs in Table \ref{disptable} have a total mass of
$\sim 1.4 \times 10^7$ \msun, comparable to the aggregate mass of the
much older Galactic globular clusters.  (However, we do not mean to
imply that all of the M82 SSCs will remain bound for 12 Gyr; see
\S\ref{discussion}.)

The old globular clusters of the Milky Way are spread widely in the
$\sigma-r_h$ space of Figure \ref{virialplot}, but in general the
locus of points is below (lower velocity dispersion and to the right
(larger radius) of the locus of points for the young M82 SSCs.  What
can we infer about the two populations from this plot?  It is
interesting to consider the time evolution of a cluster in this
parameter space.  Over time, mass loss from individual stars in the
course of their evolution will cause a cluster to lose mass.  As
detailed in the Appendix, adiabatic mass loss by a virialized cluster
progresses such that the product $\sigma r$ is conserved.  An isolated
cluster, evolving through adiabatic mass loss, would gradually move
down and to the right in Figure \ref{virialplot} as indicated by the
plotted vector, crossing the ``isobaric'' lines.  Over the span of 15
Gyr, a cluster with a Kroupa IMF would lose around half of its initial
mass (i.e., its mass at 10 Myr) as a result of stellar evolution
\citep{mccrady03}.  Such adiabatic evolution of the young M82 SSCs
over a Hubble Time would place the clusters in the same region as the
bulk of the old globular clusters in Figure \ref{virialplot}.  We
discuss the implications of this plot further in \S \ref{discussion}.

Additional context for our results comes from observed cluster systems
in other galaxies.  \citet{dubath97} measured the masses of nine
globular clusters in M31.  They find velocity dispersions of $7-27$
\kms ($\overline{\sigma} = 14$ \kms), $r_{hp} =2-5$ pc
($\overline{r_{hp}} = 3.6$ pc) and masses of $4.3-82 \times 10^5$
\msun ($\overline{M} = 2.3 \times 10^6$ \msun).  \citet{larsen02}
measure virial masses for four globular clusters in M33.  They find
velocity dispersions of $4.4-6.5$ \kms, $r_{hp}=2-8$ pc and masses of
$1.4-6.2 \times 10^5$ \msun.  In general, the old globular clusters in
these neighboring galaxies are larger with lower velocity dispersions.
This pattern is consistent with the notion that the M82 SSCs are a
population of young globular clusters, as adiabatic mass loss due to
stellar evolution would cause the clusters to expand over time (see
Appendix).  A more direct comparison is provided by young SSCs in
other galaxies.  \citet{mengel02} examines five young (age $\sim 8$
Myr) clusters in the merging Antennae galaxies.  They find velocity
dispersions of $9-21$ \kms, $r_{hp}$ of $3.6-4.0$ pc, and masses of
$6.4-47 \times 10^5$ \msun.

As discussed in \citet{mccrady05}, SSC-F shows evidence of mass
segregation.  While cluster-wide dynamical mass segregation in these
young clusters is unlikely, it is possible that the most massive stars
in a cluster have either rapidly sunk toward the core or simply formed
nearer the core.  In either case, the red supergiant velocity
dispersions we measure in the near-IR would be smaller than the
cluster mean and the masses we derive would represent lower limits.
Gasdynamical modeling by \citet{boily05} indicates that inward
migration of massive stars may cause the dimensionless geometric
parameter $\eta$ in the virial mass formula (Equation \ref{virial}) to
increase by a factor of around two over a few $\times 10^7$ yr.  In
our analysis, we have explicitly assumed that $\eta = 10$ as derived
in \citep{mccrady03}.  If mass segregation has in fact led to $\eta >
10$ in these M82 clusters, the masses we derive here would be
underestimated by the corresponding factor.

\subsection{Cluster Mass Function}

\begin{figure}
\centering 
\epsscale{1.0} 
\plotone{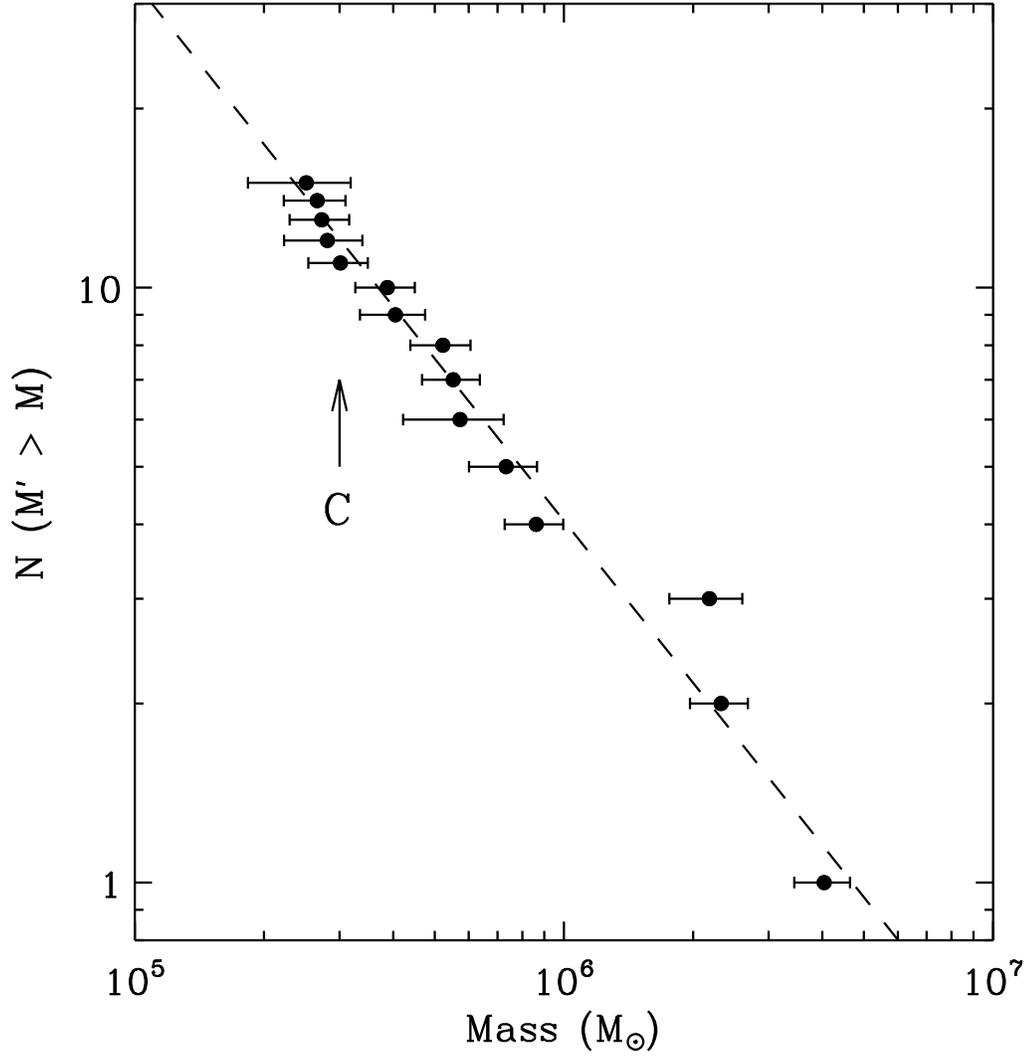}
\caption
{Cumulative mass function for the M82 SSCs.  The dashed line indicates
a power law fit where $N (M' > M) \propto M^{\gamma + 1}$.  The best
fit has a slope of $\gamma = -1.91 \pm 0.06$.  The estimated
completeness point for cluster mass is marked 'C' (see text). The
fitted power law does not reflect any correction for completeness.}
\label{massfunc}
\end{figure}

Armed with the virial masses of the SSCs, we can investigate the
cluster mass function for the M82 nuclear starburst.  The standard
manner for making a cluster mass function is to prepare a histogram of
the masses over logarithmically spaced bins and fit a power law or
lognormal distribution to the slope.  \citet{eros05} demonstrates the
shortcomings of this method, citing specifically the dependence of the
power law index to the choice of bin size and spacing in cases
involving a small number of data points.  We choose instead to
evaluate the cumulative mass function for the 15 M82 SSCs.  Figure
\ref{massfunc} plots the cumulative mass function as $N(M' > M)$,
which is the total number of clusters with mass greater than the
reference mass $M$.  In the common case of a power law, the slope of a
standard mass function is $dN/dM \propto M^{\gamma}$.  For the
cumulative mass function, the integration adds one to the exponent, such
that $N(M' > M) \propto M^{\gamma+1}$ \citep{eros05}.

The mass function for the M82 SSCs is well fit by a power law of index
$\gamma = -1.91 \pm 0.05$.  The uncertainty on the power law index is
based on a Monte Carlo simulation of our mass data.  We resampled the
cluster masses by adding normal noise according to the uncertainties
on the mass measurements, then fit for the index of the resulting
cumulative mass function.  The distribution of power law indices was
Gaussian with a standard deviation of our quoted uncertainty.  The
power law index of $\gamma \sim -2$ indicates that the stellar mass of
the cluster population is divided rather equally between the high-mass
clusters and low-mass clusters.  Of the aggregate mass in our sample
of 15 SSCs, roughly 60 percent of the mass is contained in the three
most massive clusters (SSCs L, 9 and 7).

Estimation of the completeness limit for our SSC mass function is
somewhat difficult.  Our mass measurements are based on measured
cluster velocity dispersions.  For us to measure a velocity
dispersion, the cluster must be observable in the near-IR and be
sufficiently bright and spatially resolved for us to obtain a usable
spectrum.  Limiting factors include the intrinsic mass of the cluster,
the light-to-mass ratio for the cluster, confusion with adjacent
clusters or background emission, and line-of-sight extinction.  With
the exception of SSC 1b, which suffers from source confusion, we are
confident we have obtained the spectra of all clusters brighter than
apparent magnitude $H \sim 13.8$.  At the distance of M82, this
corresponds to a cluster luminosity of $\sim 1.8 \times 10^5$ \lsun .
To convert to a mass estimate, we can estimate the light-to-mass ratio
for the clusters. For clusters in the age range of 7--13 Myr, suitable
for the M82 nuclear clusters \citep{mccrady03}, and a \citet{kroupa01}
field star mass function, Starburst99 models predict $L/M \sim 1$ in
units of \lsun/\msun.  The highly variable extinction in the dusty,
inclined disk of the galaxy could be hiding additional bright
clusters, on the far side of the disk for example.  Assuming a typical
extinction correction for the clusters of $\sim 0.5$ mag in $H$ band,
we estimate that our mass function is largely complete for clusters
more massive than $\sim 3 \times 10^5$ \msun.  To characterize the
potential effects of incompleteness, we added fake clusters of mass
$(3 \pm 0.5) \times 10^5$\msun to the Monte Carlo simulation.  Each
additional undetected cluster with mass near this completeness limit
would decrease the power law index by $\Delta \gamma \sim -0.02$
(i.e., the power law would become more steeply negative).

\section{Discussion}
\label{discussion}

In \S \ref{derivedmasses}, we note that the evolution of an individual
M82 SSC via adiabatic mass loss due to stellar evolution over a Hubble
time would reposition the SSC in $\sigma - r_h$ parameter space.  Such
a repositioning would leave any of the SSCs in our sample in a region
consistent with the position of the old globular clusters in the Milky
Way (Figure \ref{virialplot}).  In a sense, this represents a
necessary but not sufficient condition for the hypothesis that these
young SSCs represent the progenitors of globular clusters.  If,
instead, our analysis showed that stellar evolution would move the
clusters to a point in $\sigma-r_h$ parameter space that would be
inconsistent with the position of old globular clusters, it would
represent strong evidence that these SSCs could not be progenitors of
globular clusters.  In fact, adiabatically evolving any individual M82
SSC from our sample for 15 Gyr would leave it solidly within the region
of $\sigma - r_h$ parameter space occupied by old globular clusters.

But we hasten to note that this result is insufficient evidence that
these SSCs are destined to become a population of old globular
clusters.  The M82 nuclear clusters are young, with ages on the order
of 10 Myr \citep{natascha98,mccrady05th}.  There is growing evidence
in the literature that a significant portion of young clusters are
disrupted on a timescale of approximately 10 Myr from birth.
Observations of massive clusters in the Antennae galaxies
\citep{fall05} and M51 \citep{bastian05} and lower mass open clusters
in the solar neighborhood \citep{lada03} find an excess of clusters
with ages $\sim 10$ Myr relative to what would be expected based on an
assumption of constant cluster formation rate \citep{bastian06}.  The
naive interpretation of these findings is that there was a burst of
star cluster formation in the past 10 Myr in each of these galaxies.
But as noted by \citet{fall05}, the age distribution of star clusters
represents the combined histories of star cluster formation and
disruption within a galaxy.  The relative wealth of clusters aged 10
Myr in these various nearby galaxies suggests that we are fortunate to
be observing them at a special time in their star formation histories,
whereby we fall afoul of the cosmological principle.

An attractive alternative is that a high percentage of clusters are
disrupted within approximately 10 Myr of formation, i.e., the
e-folding survival time for a population of clusters is about 10 Myr
\citep{mengel05}.  This hypothesis goes by the morbid name of ``infant
mortality.''  The energy and momentum output of massive stars via
stellar winds and supernovae could remove residual natal gas from a
young massive cluster.  If the gas removal were impulsive (i.e.,
occured over less than a crossing time), the cluster could become
gravitationally unbound and begin expanding freely.  Whether or not a
cluster survives this phase depends largely on the star formation
efficiency in the formation of the cluster from natal gas \citep[see
references in][]{bastian06}.  \citet{fall05} posit that because a
cluster with more mass has both more gas to remove and more massive
stars to provide the energy, the fraction of clusters disrupted may be
roughly independent of mass.  They find this conjecture to be
consistent with their observation that the shape of the cluster mass
function in the Antennae galaxies is nearly independent of age.  

\citet{zhang99} investigated the mass function of young star clusters
in the Antennae galaxies (NGC 4038/9) based on photometric mass
estimates, and found a power law mass function of $\gamma = -2$ over
the range $10^4 \le M/\msun \le 10^6$, a result confirmed by
\citet{fall05}.  \citet{mengel05} found potential evidence for a
turnover or change in slope of the mass function for the Antennae
clusters, but cautions that the random and systematic uncertainties
discourage overinterpretation of this result.  They note that
determination of the cluster mass function from photometric masses
requires age determinations for individual clusters, which is
particularly delicate work around ages of $\sim 10^7$ yr when the
cluster luminosity varies greatly with age.

Our virial mass measurements obviate determination of the ages of
individual clusters and assumptions regarding the form and cutoff
masses of the stellar IMF.  The M82 nuclear clusters in our sample are
also young, and follow a power law mass distribution very similar to
the Antennae clusters.  A power law mass distribution for young SSCs
stands in contrast to the lognormal mass distribution for old globular
clusters \citep{harris91}, which imply a preferred mass scale at the
peak of $\sim 2 \times 10^5 \msun$ for Milky Way globular clusters.
Several processes, operating on different timescales, have the ability
to disrupt star clusters.  As discussed above, infant mortality
appears to disrupt clusters independently of mass, thereby preserving
the shape of the initial cluster mass function.  On longer timescales
($\sim 10^8-10^9$ yrs), the strongly mass dependent processes of
two-body relaxation and external perturbations (such as gravitational
shocks and dynamical friction) can disrupt the clusters
\citep{bastian05}.  Analytical models by \citet{fall01} find that the
initial form of the high-mass end of the cluster mass function is
preserved over time.  Two-body relaxation decreases the masses of
clusters linearly over time, flattening the mass function at low
masses but little affecting the shape at high masses.  By 12 Gyr, the
mass function develops a peak at a mass of about $2 \times 10^5
\msun$.  Thus, an initial power law distribution of cluster masses
will develop into a distribution resembling the lognormal mass
function of old globular clusters over 12 Gyr, with disruption erasing
all information regarding the initial shape of the mass function at
the low-mass end.

In a galactic environment, of course, the clusters are not isolated,
and are additionally subject to external forces such as galactic tides
and dynamical friction.  The inner 1 kpc of a galaxy, with its strong
tidal fields, is a particularly dangerous place for star clusters.
But as noted by \citet{fall01}, their models ``support the suggestion
that at least some of the star clusters formed in merging and
interacting galaxies can be regarded as young globular clusters.''  As
shown in Figure \ref{virialplot}, the adiabatic evolution of a cluster
through stellar evolution and mass loss over a similar time frame will
tend to move the M82 SSCs into the $\sigma-r_h$ parameter space
occupied by old Galactic globular clusters.  While we cannot predict
the fate of any individual M82 SSC, our results suggest that any
cluster which should happen to survive for a Hubble time could
resemble the old globular clusters seen in the Milky Way today.

\section{Summary}

In this paper, we investigate the SSC population of the inner $\sim
500$ pc of the M82 starburst.  The nuclear starburst in M82 contains
roughly two dozen SSCs that are prominent in the near-IR.  Based on
high spectral resolution near-IR spectra, we measure line-of-sight
velocity dispersions for 19 SSCs in the nuclear starburst.  We find
dispersions in the range of $7-35$ \kms , comparable with values for
older globular clusters.  We apply the virial theorem to the measured
velocity dispersions and halflight radii to derive the masses of 15 of
the SSCs.  The SSC masses lie in the range of $2.5 \times 10^5$ \msun
to $4 \times 10^6$ \msun, placing them at the high end of the mass
distribution function for old Galactic globular clusters.  The total
mass of the 15 measured SSCs is $1.4 \times 10^7$ \msun, which is of
the same order of magnitude as the total mass of the globular cluster
system of the Milky Way.  Evolution of the clusters via gradual mass
loss from stellar evolution would move them into the realm of
$\sigma-r_h$ parameter space occupied by old Milky Way globular
clusters.  The cumulative mass function of the clusters follows a
power law with an index of $\gamma = -1.91 \pm 0.06$.  This is very
similar to the mass distribution of young SSCs in the Antennae
galaxies, and lends credence to the suspicion that SSCs are potential
future globular clusters.

\acknowledgments

We would like to thank the staff of the Keck Observatory for their
assistance in our observations.  We also thank the anonymous referee
for helpful comments regarding implications of this work.  NM thanks
John Johnson for invaluable data wrangling advice and W.~D.~Vacca,
L.~Blitz and S.~E.~Boggs for helpful comments.  The authors wish to
recognize and acknowledge the very significant cultural role and
reverence that the summit of Mauna Kea has always had within the
indigenous Hawaiian community.  We are most fortunate to have the
opportunity to conduct observations from this mountain.  This material
is based upon work supported by the National Science Foundation under
Grant No. 0502649, with additional support from NSF Grant
AST--0205999.  Any opinions, findings, and conclusions or
recommendations expressed in this material are those of the authors
and do not necessarily reflect the views of the National Science
Foundation.

\appendix
\section{Appendix}
\label{appendix}

The consequences of mass loss from a virialized star cluster depend
upon the rate of loss.  We consider specifically two cases: (1) rapid
mass loss, where a star cluster has a sufficiently long relaxation
time that $v$ and $R$ are unable to readjust during the ejection of
some quantity of mass, and (2) slow mass loss, where $v$ and $R$ for
the cluster continually readjust to maintain equilibrium.

In the case of rapid mass loss, \citet{hills80} derives an expression
for the relation between the initial mass and the amount of mass
ejected:
\begin{equation}
R = R_0 \frac{M_0 - \Delta M}{M_0 - 2 \Delta M}.
\end{equation}
Evidently, as $\Delta M$ tends to $ M_0/2$ the cluster radius tends to 
infinity and the cluster becomes unbound.

In the case of gradual mass loss, the cluster constantly adjusts to
maintain equilibrium.  The total energy at any instant, whether the
cluster is in equilibrium or not, is
\begin{equation}
E = \frac{1}{2}Mv^2 - \eta \frac{GM^2}{R}
\end{equation}
where $v$ is the 3-d rms velocity and $\eta$ is a non-dimensional form
factor.  The corresponding change in total energy of the system is
\begin{equation}
\left(\frac{\partial E}{ \partial M} \right)_{R,v} =
\frac{1}{2}v^2 - 2\eta \frac{GM}{R}. 
\end{equation}
After $\delta m$ is lost, the cluster must readjust to the new
equilibrium. In the new equilibrium configuration the Virial theorem,
$E=-T=\Omega/2$ \citep[][p. 211]{binney87}, can be invoked to express
the partial derivative:
\begin{equation}
\left(\frac{\partial E}{ \partial M} \right)_{R,v} =
-\frac{E}{M} + 4\frac{E}{M} =  3\frac{E}{M},
\end{equation}
which can be integrated from the
initial mass, $M_0$, and energy $E_0$, 
\begin{equation}
\int_{E_0}^E dE/E = 3\int_{M_0}^M dM/M
\end{equation}
to yield $E/E_0 = (M/M_0)^3$. Substituting again from the Virial
theorem, which shows that $E \propto \Omega \propto M^2/R$, we have
have an expression for the radius as a function of mass,
\begin{equation}
R = R_0 \left(\frac{M}{M_0}\right)^{-1}.
\label{adiabatic}
\end{equation}
which is the adiabatic invariant from \citet{hills80},
where $M \equiv M_0 - \Delta M$.

In the equilibrium states both before and after loss of $\Delta M$,
the Virial theorem applies, and $E=-T$ (where $T=Mv^2/2$ is the
kinetic energy).  Thus:
\begin{equation}
\frac{E}{E_0} = \left(\frac{M}{M_0}\right)^3 = \frac{-Mv^2/2}{-M_0 v_0^2/2}\,.
\end{equation}
Simplifying and combining terms,
\begin{equation}
\left(\frac{M}{M_0}\right)^2 = \left(\frac{v}{v_0}\right)^2
\end{equation}
which has the physical solution $(M/M_0) = (v/v_0)$.  If we compare
this last result to Equation (\ref{adiabatic}), we find
\begin{equation}
\frac{M}{M_0} = \frac{v}{v_0} = \frac{R_0}{R} 
\end{equation}
and therefore for adiabatic mass loss, $vR = v_0 R_0 =$ constant.


\bibliographystyle{apj}
\bibliography{apj-jour,astro_refs}



\begin{deluxetable}{lcccccc}
\tablewidth{0pt}
\tablecolumns{7}
\tablecaption{NIRSPEC N5 Observations}
\tablehead{\colhead{Date} & \colhead{Objects} & \colhead{$t_{exp}$} & 
    \colhead{Airmass} & \colhead{Seeing} & \colhead{Atm Star} &
    \colhead{Remarks}\\
    \colhead{(UT)} & \colhead{} & \colhead{(min)} & 
    \colhead{(sec $z$)} & \colhead{($''$)} & \colhead{(SpT)} &
    \colhead{}}
\startdata
2002 Feb 23  & 3, 9, 11  &  40 & 1.6 & 0.5  & HD 74604 (B8V) & (1) \\
2003 Jan 19  & F, L      &  70 & 1.9 & 0.8  & HD 173087 (B5V) & \\
2003 Feb 6   & 1b, 1c, r &  60 & 1.8 & 0.7  & HD 74604 (B8V) & \\
2003 Feb 6   & 1a, 3, m  &  50 & 1.7 & 0.7  & HD 82327 (B9V) & (2)\\
2003 Feb 6   & 6, 7      &  60 & 1.6 & 0.7  & HD 82327 (B9V) & \\
2003 Feb 6   & 8, 10, c  & 120 & 1.6 & 0.6  & HD 82327 (B9V) & \\
2003 Feb 6   & s, t      &  60 & 1.9 & 0.6  & HD 146926 (B8V) & \\
2003 Feb 7   & a, b      &  90 & 2.0 & 1.5  & HD 146926 (B8V) & (3)\\
2003 Dec 5   & 6, 7      &  20 & 1.6 & 0.5  & HD 63586 (A0V) & \\
2003 Dec 5   & 3, 6, h   &  40 & 1.6 & 0.5  & HD 63586 (A0V) & (4) \\
2004 Feb 8   & a, y      &  60 & 1.9 & 0.8  & HD 146926 (B8V) & \\
2004 Feb 8   & r, z      &  30 & 2.2 & 1.0  & HD 146926 (B8V) & \\
2004 Feb 9   & r, z      &  40 & 1.6 & 1.3+ & HD 82327 (B9V)  & (3) \\
2004 Feb 9   & a, y      &  30 & 1.7 & 1.3+ & HD 82327 (B9V)  & (3) \\
2005 Jan 24  & 1a, 1b    &  20 & 1.6 & 0.6  & HD 82327 (B9V)  & \\
2005 Jan 24  & 1a, 1c, q &  20 & 1.5 & 0.6  & HD 82327 (B9V)  & \\
2005 Jan 24  & j, k, q   &  30 & 1.5 & 0.6  & HD 82327 (B9V)  & \\
\enddata
\tablecomments{Seeing values are estimates.  Remarks. --- (1)~Only 30 min 
 on object 3.  (2)~Only 40 min on object 1a.  (3)~Seeing very poor \& 
variable. (4)~Only 20 min on object 3.}
\label{obs1}
\end{deluxetable}


\begin{deluxetable}{lccccc}
\tablewidth{0pt}
\tablecolumns{6}
\tablecaption{Cross-Correlation Results}
\tablehead{ & \colhead{Best Fit} & \colhead{Mean CCF} &
\colhead{Dispersion} & \colhead{Mass\tablenotemark{$\ast$}} &
\colhead{$t_{cr}$\tablenotemark{$\dagger$}} \\
\colhead{Object} & \colhead{Template} & \colhead{Peak Value} & 
\colhead{($\sigma$, \kms}& \colhead{($10^5$ \msun)} & \colhead{($10^5$ yr)} 
}
\startdata
SSC L & HR2289 & $0.70 \pm 0.05$ & $34.7 \pm 0.4$ & $40. \pm 6.$ & $0.41 \pm 0.05$ \\
SSC F & HR2289 & $0.69 \pm 0.1$ & $12.4 \pm 0.3$ & $5.5 \pm 0.8$ & $1.2 \pm 0.1$ \\
SSC a & HD237008 & $0.53 \pm 0.1$ & $10.9 \pm 0.4$ \\
SSC 11 & HD237008 & $0.78 \pm 0.07$ & $12.1 \pm 0.4$ & $3.9 \pm 0.6$ & $0.9 \pm 0.1$ \\
SSC 9 & HD237008 & $0.70 \pm 0.07$ & $19.8 \pm 0.5$ & $23. \pm 4.$   & $1.3 \pm 0.2$  \\
SSC 8 & HD14469 & $0.52 \pm 0.1$ & $10.5 \pm 0.5$ & $4.0 \pm 0.7$    & $1.5 \pm 0.2$  \\
SSC 7 & HD237008 & $0.53 \pm 0.1$ & $18.6 \pm 1.$ & $22. \pm 4.$     & $1.4 \pm 0.2$  \\
SSC 6 & HD237008 & $0.76 \pm 0.09$ & $9.2 \pm 0.3$ & $2.7 \pm 0.4$  & $1.5 \pm 0.2$  \\
SSC h & HD237008 & $0.62 \pm 0.07$ & $33.2 \pm 1.0$ \\
SSC j & HD237008 & $0.55 \pm 0.10$ & $9.0 \pm 0.8$ \\
SSC k & HD237008 & $0.51 \pm 0.1$ & $9. \pm 1.$ & $5.7 \pm 2.$     & $3.1 \pm 0.5$  \\
SSC m & HD14469 & $0.38 \pm 0.1$ & $15.2 \pm 0.8$ & $7.3 \pm 1.$     & $0.9 \pm 0.1$ \\
SSC q & HD237008 & $0.53 \pm 0.1$ & $7.9 \pm 0.6$ & $2.8 \pm 0.6$   & $2.5 \pm 0.3$  \\
SSC 3 & HD237008 & $0.76 \pm 0.1$ & $8.7 \pm 0.3$ & $2.7 \pm 0.4$   & $1.8 \pm 0.2$  \\
SSC 1a & HD237008 & $0.72 \pm 0.07$ & $13.4 \pm 0.4$ & $8.6 \pm 1.$  & $1.5 \pm 0.2$  \\
SSC 1c & HD237008 & $0.71 \pm 0.08$ & $12.2 \pm 0.4$ & $5.2 \pm 0.8$ & $1.2 \pm 0.2$  \\
SSC r & HD237008 & $0.65 \pm 0.1$ & $8.6 \pm 0.3$ & $3.0 \pm 0.5$   & $2.0 \pm 0.2$  \\
SSC t & HD237008 & $0.46 \pm 0.1$ & $7.9 \pm 0.9$ & $2.5 \pm 0.7$   & $2.2 \pm 0.4$  \\
SSC z & HD237008 & $0.69 \pm 0.1$ & $9.9 \pm 0.3$ \\
\enddata
\tablenotetext{$\ast$}{Error in mass includes errors in distance
to M82, half-light radius and velocity dispersion.}
\tablenotetext{$\dagger$}{Crossing time, described
in \S \ref{derivedmasses}.}
\label{disptable}
\end{deluxetable}

\clearpage



\clearpage

\clearpage


\end{document}